\documentclass[epj]{svjour}
\usepackage{graphics}
\usepackage{authblk}
\usepackage[utf8]{inputenc}
\usepackage[english]{babel}
\usepackage{indentfirst}
\usepackage{cite}
\usepackage{graphicx}
\usepackage{amsmath,amsfonts}
\usepackage{amssymb}
\usepackage{float}
\begin{document}
\title{Coupled Lugiato-Lefever equation for nonlinear frequency comb generation at an avoided crossing of a microresonator}
\author{Giuseppe D'Aguanno\inst{1}\thanks{\emph{Corresponding author:} giusdag@utexas.edu or giuseppe.daguanno@gmail.com}\and Curtis R. Menyuk\inst{2}
%
}                     
%
%
\institute{Department of Electrical and Computer Engineering,
The University of Texas at Austin, Austin, Texas, 78712, USA \and Department of Computer Science and Electrical Engineering, University of Maryland, 1000 Hilltop Circle, Baltimore, Maryland
21250, USA}
\date{Received: date / Revised version: date}
%
\abstract{
Guided-mode coupling in a microresonator generally manifests itself through avoided crossings of the corresponding resonances. This coupling  can strongly modify the resonator local effective dispersion by creating two branches that have dispersions of opposite sign in spectral regions that would otherwise be characterized by either positive (normal)  or negative (anomalous) dispersion.  In this paper, we study, both analytically and computationally, the general properties of nonlinear frequency comb generation at an avoided crossing using the coupled Lugiato-Lefever equation. In particular, we find that bright solitons and broadband frequency combs can be excited when both branches are pumped for a suitable choice of the pump powers and the detuning parameters. A deterministic path for soliton generation is found.
} 
\maketitle
\section{Introduction}
\label{intro}
\begin{figure}
\resizebox{0.5\textwidth}{!}{%
  \includegraphics {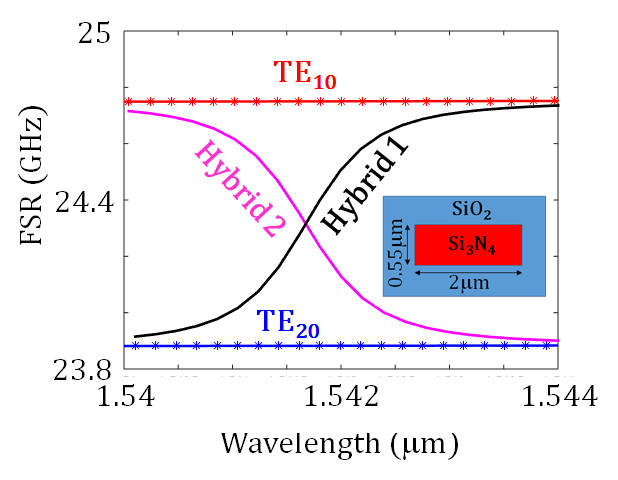}
}
\caption{ Shown is an example of a simulated avoided crossing for a  $\rm{Si_{3}N_{4}}$ resonator embedded in  $\rm{SiO_{2}}$ with radius $942.8~{\rm{\mu m}}$ (total path length 5.92 mm) and $2~{\rm{\mu m}} \times 550~\rm{nm} $ waveguide cross-section. The inset on the right is a schematic drawing of the waveguide cross-section. The details of the figure are described in the main text.}
\label{Avoided_Crossing}       
\end{figure}

\indent  High-quality (Q) microcavities play a fundamental role in linear and nonlinear optics \cite {Vahala,Notomi}. The strength of the nonlinear light-matter interactions in a microcavity 
is proportional to the product of the atomic transition rate of the medium filling the cavity and
the cavity Q-factor. Hence, a high  Q increases the efficiency of the nonlinear interaction and thus enhances the material's effective nonlinearity \cite {Vahala}. Among high-Q optical microcavities, a particularly interesting category is the one that includes dielectric micro-toroid resonators \cite {Armani} and dielectric microspheres \cite {Braginsky,Birks}. In both the micro-toroid resonator and the microsphere, the electromagnetic field is concentrated in the immediate vicinity of the dielectric-air interface and propagates along the circumference in a similar fashion to what acoustic waves do in a circular acoustic resonator, as first described by Lord Rayleigh \cite{Rayleigh} for the whispering gallery of St.~Paul's Cathedral in London, from which the name ``whispering-gallery-mode" (WGM) resonators is derived. 
In the last few years a great deal of theoretical and experimental effort has been devoted to the study 
of mode-locked soliton generation in WGM resonators with a Kerr nonlinearity \cite {DelHaye,Matsko, Herr, Coen, Chembo,Godey}. In the Fourier domain, the mode-locked train of solitons gives rise to a frequency comb made of narrow and nearly equidistant spectral lines, which is promising for applications  to metrology,
high-resolution spectroscopy, and microwave photonics.  
 In particular, recent theoretical and experimental efforts \cite {Brasch,Matsko1} have demonstrated that the spectrum of a soliton train generated in the anomalous dispersion region of the resonator can be broadened well into the normal dispersion region due to the mode interaction and the  emission of soliton Cherenkov radiation. Achieving a broadband frequency comb with more than one octave of bandwidth in a microresonator is important because   it allows the measurement of  the carrier-envelope offset \cite {Diddams} and might open the door to self-referencing combs on chip-scale dimensions \cite{Yi, Li1}. Other experiments have also shown that strong modification of the effective dispersion properties of the resonator with respect to the material properties can occur in spectral regions near the avoided-mode-crossing points of the resonator \cite {Gaeta,Weiner}. In these regions, two frequency-degenerate guided modes of the resonator undergo a strong linear interaction, with a group velocity mismatch (GVM)  practically equal to zero, leading to the formation of two new hybrid guided modes with group velocity dispersions (GVDs) of opposite sign. 
An example is provided in Fig.~\ref{Avoided_Crossing}. In the  example an avoided crossing is simulated for a  $\rm{Si_{3}N_{4}}$ resonator embedded in  $\rm{SiO_{2}}$ with radius $942.8~{\rm{\mu m}}$ (total path length 5.92 mm) and a waveguide cross-section $2~{\rm{\mu m}} \times 550~\rm{nm} $. The corresponding waveguide admits two transverse electric (TE) guided modes, $\rm{TE_{10}}$ and $\rm{TE_{20}}$, in the normal dispersion regime. A family of resonator eigenfrequencies, indicated with asterisks in the figure, can be calculated for each one of the guided modes. Close to the wavelength of $\lambda=1.542~{\rm{\mu m}}$, the resonator eigenfrequency of the  $\rm{TE_{10}}$ mode and that of the $\rm{TE_{20}}$ mode are degenerate.  The two frequency-degenerate guided modes of the resonator undergo a strong linear interaction with a group velocity mismatch practically equal to zero. This strong interaction leads to the formation of two new hybrid guided modes ($\omega_{\pm}$), that we call respectively Hybrid 1 and Hybrid 2 in the figure, that are no longer frequency-degenerate, and whose frequency splitting depends on the  coupling strength of the frequency-degenerate modes. This resonant mode-coupling is described by the following equation
\begin {equation}\normalsize
\omega_{\pm}=\frac{\omega_{\rm{TE_{10}}}+\omega_{\rm{TE_{20}}}}{2}\pm\sqrt{\frac{(\omega_{\rm{TE_{10}}}-\omega_{\rm{TE_{20}}})^2}{4}+\kappa^2}
\tag {1}
 \end {equation}
where $\omega_{\rm{TE_{10}}}$ is the resonator eigenfrequency of the $\rm{TE_{10}}$ mode, $\omega_{\rm{TE_{20}}}$ that of the $\rm{TE_{20}}$ and $\kappa$ is the coupling coefficient. The value of the coupling coefficient has been chosen according to the experimental data reported in \cite {Weiner}.
  In this case, one of the two modes (Hybrid 2) acquires anomalous dispersion in a spectral region that would otherwise be characterized by normal dispersion. The figure shows the free spectral range (FSR) vs. wavelength of the unperturbed modes, $\rm{TE_{10}}$ and $\rm{TE_{20}}$, and that of the perturbed modes. It is noted that both unperturbed modes have normal dispersion, {\it{i.e.}} their FSR increases  as the wavelength increases, while the dispersion is normal for Hybrid 1 and anomalous for Hybrid 2.\\
\indent In a recent publication \cite {D'Aguanno}, the coupled Lugiato-Lefever equation (CLLE) that governs  the  nonlinear interaction between two families of  modes with different transverse profiles in a generic WGM resonator was derived from first principles. In this paper,  we use the CLLE to study the general properties of nonlinear frequency comb generation at an avoided crossing. In particular, we find that bright solitons and broadband frequency combs can be excited when both branches are indipendently pumped for a suitable choice of the pump powers and the detuning parameters. Moreover, the soliton generation follows a deterministic path. The use of two independent pumps plays a critical role in the deterministic generation of solitons. While use of independent pumps has previously been discussed \cite{Gaeta1, Wabnitz}, their usefulness in the deterministic generation of solitons has not been pointed out previously.\\ 
Nonlinear interactions are ubiquitous and the phenomenon of avoiding crossings does not only appear in guided-wave optics, but also appears in many other fields such as quantum chemistry, nuclear physics, quantum electrodynamics and  quantum chromodynamics \cite {Heiss}. Hence, we expect that the results discussed in this paper will have a similar broad range of applicability.
 
\section{Results and Discussion}
\indent We start with the CLLE, recently derived in Ref.~\cite{D'Aguanno}, that governs the nonlinear interaction of two modes in a generic WGM resonator:
\begin {equation}\normalsize
\frac{\partial \psi^{(j)}}{\partial \tau}=\delta^{(j)}\frac{\partial \psi^{(j)}}{\partial \theta}-i\frac{\bar\beta_2^{(j)}}{2}\frac{\partial^2 \psi^{(j)}}{\partial \theta^2}-\left(\frac{\bar\tau}{\tau^{(j)}}+i\alpha^{(j)}\right)\psi^{(j)}
\notag
 \end {equation}
\begin {equation}\normalsize
+ih^{(j)}+i\psi^{(j)}\sum\limits_{k=1}^{2}D^{(j,k)}|\psi^{(k)}|^2,\hspace{11mm}
\tag{2}
\end {equation}
with $j=1,2$. In Eq.~(2), $\tau=t/\bar\tau$ is the time normalized to the average cavity photon lifetime of the two modes $[\bar\tau=(\tau^{(1)}+\tau^{(2)})/2]$; $\theta$ is the resonator azimuthal coordinate in the retarded coordinate system;  $\delta^{(j)}$ is the normalized GVM; $\alpha^{(j)}=\delta\omega^{(j)}\bar\tau$ is the normalized detuning;  $\delta\omega^{(j)}=\omega_{\bar m}^{(j)}-\omega_p^{(j)}$ is the detuning of the frequency of the pump field with respect to the $\bar m^{th}$ cavity eigenfrequency; $\bar m$ labels the cavity eigenfrequency closest to the pump frequency, which we call the dominant eigenfrequency; $\omega_p^{(j)}$ is the pump frequency associated with the $j^{th}$ mode ($\omega_p^{(1)}\cong\omega_p^{(2)}\cong\omega_p$); $\omega_{m}^{(j)}$ denotes the $m^{th}$ cavity eigenfrequency associated with the $j^{th}$ mode;  $\psi^{(j)}=\sqrt{2\chi^{(3)}Q}e^{-i\alpha^{(j)\tau}}\Psi^{(j)}$ is the dimensionless field envelope and $\Psi^{(j)}$ the field envelope; $\chi^{(3)}$ is the resonator cubic nonlinearity;  $Q\cong\bar\tau\omega_p/2$ is the cavity $Q$-factor referred to the average cavity photon lifetime; $\bar\beta_2^{(j)}=\beta_2^{(j)}\bar\tau$  is the normalized GVD parameter; $\beta_2^{(j)}=-(\omega_{\bar m+1}^{(j)}-2\omega_{\bar m}^{(j)}+\omega_{\bar m-1}^{(j)})$ denotes the GVD parameter (the dispersion is normal when $\beta_2^{(j)}>0$ and anomalous when $\beta_2^{(j)}<0$); $h^{(j)}=H^{(j)}\sqrt{2\chi^{(3)}Q^3}$ is the dimensionless pump field coupled with the $j^{th}$ mode and $H^{(j)}$ the pump field; and  $D^{(j,k)}$ are the overlap integrals of the modes. In particular, $D^{(1,1)}$ and $D^{(2,2)}$ account for the nonlinear self-coupling of mode-1 and mode-2, respectively, while $D^{(1,2)}$ and $D^{(2,1)}$ account for the cross-coupling between the two modes ($D^{(1,2)}=D^{(2,1)}$). For a detailed derivation of Eq.~(2), the reader can consult Ref.~\cite{D'Aguanno}. In general, each mode, $j=1,2$, can be independently pumped \cite {Gaeta1,Wabnitz}.  We will show that this extra degree-of-freedom makes it possible to generate a soliton deterministically.\\
\indent In the case considered here, the two modes are the two hybrid modes of an avoided crossing and Eq.~(2) can be simplified. First, at the avoided crossing, the GVM of the two interacting modes is practically zero, {\it{i.e.}} the two modes have approximately the same FSR, as shown in Fig.~1. Hence, we can set $\delta^{(j)}=0$. Second, the GVD parameters of the two modes are opposite in sign, but approximately equal in absolute value ($|\bar\beta_2^{(1)}|\cong|\bar\beta_2^{(2)}|\cong|\bar\beta_2|$). Third, the cross-coupling  terms are twice as large as the self-coupling terms.  These simplifications lead to the following 
equation   
 \begin {equation}\normalsize
i\frac{\partial U^{(1)}}{\partial \tau}-\frac{1}{2}\frac{\partial^2 U^{(1)}}{\partial \sigma^2}+(i-\alpha^{(1)}) U^{(1)}
\notag
  \end {equation}
\begin {equation}\normalsize
+U^{(1)}(|U^{(1)}|^{2}+2|U^{(2)}|^{2})=-\sqrt{P^{(1)}}\:,
 \tag{3.a}
\end {equation}
\begin {equation}\normalsize
i\frac{\partial U^{(2)}}{\partial \tau}+\frac{1}{2}\frac{\partial^2 U^{(2)}}{\partial \sigma^2}+(i-\alpha^{(2)}) U^{(2)}
 \notag
\end {equation}
\begin {equation}\normalsize
+U^{(2)}(2|U^{(1)}|^{2}+|U^{(2)}|^{2})=-\sqrt{P^{(2)}}\:,
\tag{3.b}
\end {equation}
where, without loss of generality, following the example provided in Fig.~1, we have supposed that the hybrid mode 1 is in the normal dispersion regime  and the hybrid mode 2 is in the anomalous dispersion regime. In Eq.~(3), we rescale the azimuthal coordinate, the field envelope, the pump power so that $\sigma=\theta/{(|\bar\beta_{2}|)^{1/2}}$, $U^{(j)}=\sqrt{D^{(1,1)}}\psi^{(j)}$, and $P^{(j)}=D^{(1,1)}h^{(j)2}>0$. In Eq.~(3), we also assumed that the  cavity photon lifetimes associated with the two modes are equal. If the detuning, the loss and the pump are all suppressed in Eq.~(3), Eq.~(3) becomes formally identical to the equations that describe the nonlinear, incoherent coupling of two light pulses co-propagating in a single-mode fiber in the normal and anomalous dispersion regime, respectively \cite {Trillo}.
\indent We find by substitution into Eq.~(3) that a particular class of  continuous-wave (CW) solutions is the one given by
\begin {equation}\normalsize
U_{0}^{(1)}=i\sqrt{P^{(1)}} , \:\: U_{0}^{(2)}=i\sqrt{P^{(2)}}
 \tag{4}
 \:,
 \end {equation}
with $\alpha^{(1)}=P^{(1)}+2P^{(2)}$ and $\alpha^{(2)}=2P^{(1)}+P^{(2)}$. The formation of a train of solitons is generally initiated by the modulational instability (MI) of the CW solutions \cite {Zakharov,Hasegawa}.  Hence, we search for parameters for which the MI appears. We write the field envelope as $U^{(j)}=[U_{0}^{(j)}+v^{(j)}+iw^{(j)}]$, where $v^{(j)}(\sigma,\tau)$ and $w^{(j)}(\sigma,\tau)$ are small perturbations, and we next  linearize Eq.~(3) around $U_{0}^{(j)}$. We then search for forward-propagating wave solutions in the form $v^{(j)}={\rm{Re}}\{x^{(j)}\exp[i(K \sigma-{\rm{\Omega}} \tau)]\}$ and $w^{(j)}={\rm{Re}}\{y^{(j)}\exp[i(K \sigma-{\rm{\Omega}} \tau)]\}$, where $\rm{\Omega}$ is the frequency shift with respect to the dominant frequency and $K$ is the corresponding shift in the wavenumber. Substituting the traveling wave solutions into the linearized system, we obtain the following system of linear, homogeneous, algebraic equations

\begin {equation}\normalsize
\begin{pmatrix}K^2/2&i{\rm{\Omega}}-1&0&0\\
-i{\rm{\Omega}}+1&K^2/2&0&4\sqrt{P^{(1)}P^{(2)}}\\0&0&-K^2/2&i{\rm{\Omega}}-1\\
0&4\sqrt{P^{(1)}P^{(2)}}&-i{\rm{\Omega}}+1&-K^2/2\end{pmatrix}
\begin{pmatrix}x^{(1)}\\y^{(1)}\\x^{(2)}\\y^{(2)}\end{pmatrix}=
\notag
\end{equation}
\begin {equation}\normalsize
\begin{pmatrix}0\\0\\0\\0\end{pmatrix}\:.
\tag{5}
\end{equation}
Equation (5) admits non-trivial solutions when the determinant of the matrix is zero. This compatibility condition yields  the dispersion relation, ${\rm{\Omega}}(K)$, which has four solutions
\begin {equation}\normalsize
i{\rm{\Omega}}_{(\pm,\pm)}=1\pm \left\{{-\frac{K^{4}}{4}\pm2iK^2\sqrt{P^{(1)}P^{(2)}}}\right\}^{1/2}
 \tag{\normalsize{6}}
 \:.
 \end {equation}
 The MI occurs at those values of $K$ at which  ${\rm{Re}}{(i{\rm{\Omega}})}<0$ and the traveling waves grow exponentially. The solutions $i{\rm{\Omega}}_{(+,+)}$ and $i{\rm{\Omega}}_{(+,-)}$ never satisfy the condition ${\rm{Re}}{(i{\rm{\Omega}})}<0$. Instead, the solutions $i{\rm{\Omega}}_{(-,+)}$ and $i{\rm{\Omega}}_{(-,-)}$ both satisfy this condition, but $i{\rm{\Omega}}_{(-,+)}$ corresponds to the backward propagating wave because $\rm{Im}(i{\rm{\Omega}}_{(-,+)})<0$. Hence, the solution that yields MI for the forward-propagating wave is the fourth one, $i{\rm{\Omega}}_{(-,-)}$.  Due to the $2\pi$-periodicity of the system in the azimuthal coordinate $\theta$, the wavenumber $K$ can only assume discrete values, $K=p{(|\bar\beta_2|)^{1/2}}$, where $p=m-\bar m=\pm1,\pm2,...$ is the shift of the eigenfrequency number of the perturbation with respect to the eigenfrequency number $\bar m$ of the dominant eigenfrequency. 
Explicit expressions for the eigenvector components are
\begin {equation}
x^{(1)}=C\:,
\tag{7.a}
\end {equation}
\begin {equation}
y^{(1)}=-CK^2/\Delta\:, 
\tag{7.b}
\end {equation}
\begin {equation}
x^{(2)}=\frac{C\Delta[K^4+\Delta^2]}{8K^2\sqrt{P^{(1)}P^{(2)}}\Delta}\:,
\tag{7.c}
\end {equation}
\begin {equation}
y^{(2)}=\frac{C[K^4+\Delta^2]}{8\sqrt{P^{(1)}P^{(2)}}\Delta}\:,
\tag{7.d}
\end {equation}
where $\Delta=2[i{\rm{\Omega}}_{(-,-)}-1]$ and $C$ is an arbitrary constant that quantifies the magnitude of the modulation around the CW solutions.  \\
\indent To verify the results of our analytical study, we have performed a numerical integration of Eq.~(3), using a symmetrized fast Fourier transform, split-step algorithm \cite{Fleck} with the initial conditions
\begin {equation}\normalsize
U^{(j)}(\sigma,\tau=0)=U_{0}^{(j)}+{\rm{Re}}\left[x^{(j)}\exp\left(i K\sigma\right)\right]
\notag 
\end {equation}
\begin {equation}\normalsize
\vspace{12mm}
+i{\rm{Re}}\left[y^{(j)}\exp\left(i K\sigma\right)\right] \:,
\tag{8}
\end {equation}
with $j=1,2$, where $U_{0}^{(j)}$ is given by Eq.~(3) and $x^{(j)}$ and $y^{(j)}$ are given by Eq.~(7).
The initial conditions described in Eq.~(8) are the CW solutions modulated by the solutions of the linearized system, setting $C=0.1$. In Fig.~2, we show the results of a  large scale numerical integration of Eq.~(3) in the parameter space $(P^{(1)},P^{(2)})$ to search for bright solitons at the avoided crossing. 
\begin{figure}[H]
\resizebox{0.5\textwidth}{!}{%
  \includegraphics {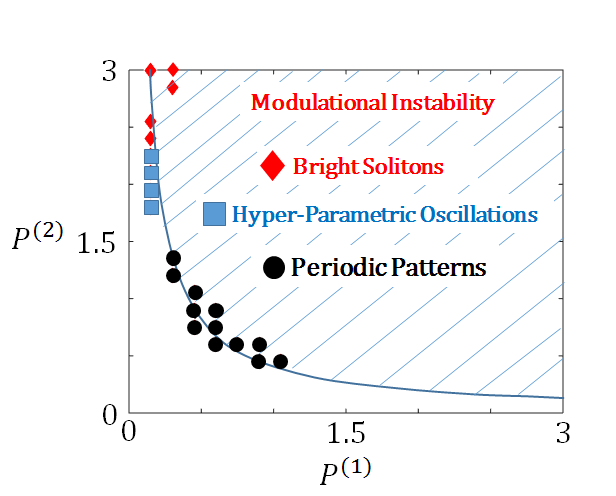}
}
\caption{ Parameter values (shaded) in the parameter space for which the MI occurs according to the dispersion relation, Eq.~(6). In this case, the dispersion has been calculated at $K=16(|\bar\beta_2|)^{1/2}$ for the fourth solution, $i{\rm{\Omega}}_{(-,-)}$.  The circles, the squares and the diamonds  indicate the position of spatio-temporal periodic patterns, hyperparametric oscillations and bright solitons, respectively. The results have been  obtained by the numerical integration of Eq.~(3) with the initial conditions reported in Eq.~(8). The numerical integration has been performed for the following values of the parameters: $\alpha^{(1)}=P^{(1)}+2P^{(2)}$, $\alpha^{(2)}=2P^{(1)}+P^{(2)}$, $K=16(|\bar\beta_2|)^{1/2}$, and $|\bar\beta_2|=0.01$. It is noted that all the states exist near the border of the MI region, as expected.}
\end{figure} 
The numerical integration has been performed using the following values of the remaining parameters: $\alpha^{(1)}=P^{(1)}+2P^{(2)}$, $\alpha^{(2)}=2P^{(1)}+P^{(2)}$, $K=16(|\bar\beta_2|)^{1/2}$, and $|\bar\beta_2|=0.01$. Three different regimes can be clearly identified near the border of the MI region. The first regime is obtained when both branches are pumped with comparable powers ($P^{(1)}\approx P^{(2)}$). In this case, the light self-organizes into spatio-temporal periodic patterns. An example is provided in Fig.~3.
\begin{figure*}
\begin {center}
\resizebox{1\textwidth}{!}{%
  \includegraphics {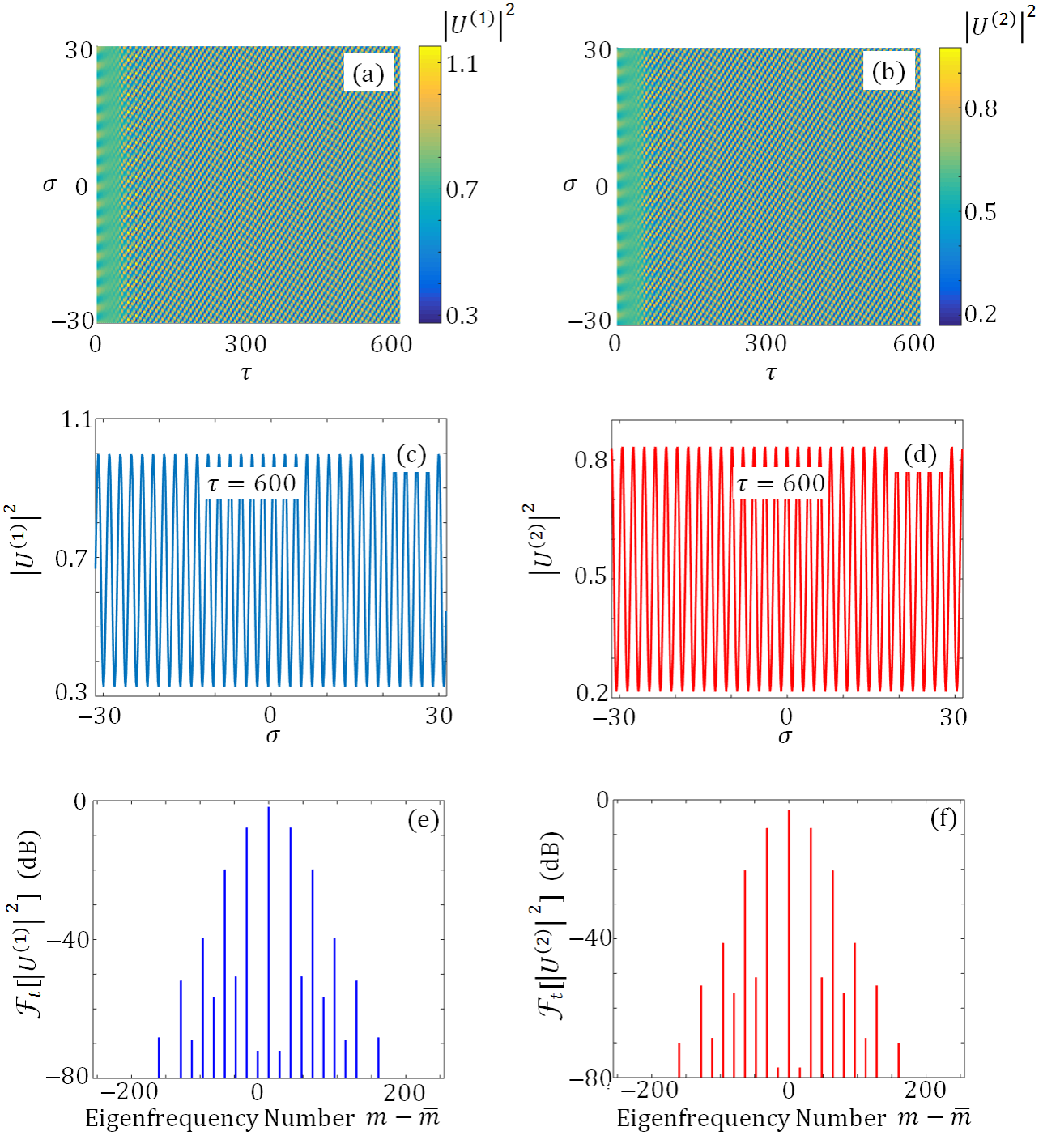}}
\end {center}
\caption{Example of a spatio-temporal periodic pattern calculated for $P^{(1)}=0.75$ and $P^{(2)}=0.6$. (a) and (b): Spatio-temporal evolution of the numerical solutions. (c) and (d): Numerical solutions calculated at $\tau=600$. (e) and (f): Fourier transform (${\mathcal{F}}_{t}$) of the solutions.}
\end{figure*}
\indent By increasing the pumping level on the anomalous dispersion branch, hyperparametric oscillations are obtained. An example is provided in Fig.~4. 
\begin{figure*}
\begin {center}
\resizebox{1\textwidth}{!}{%
  \includegraphics {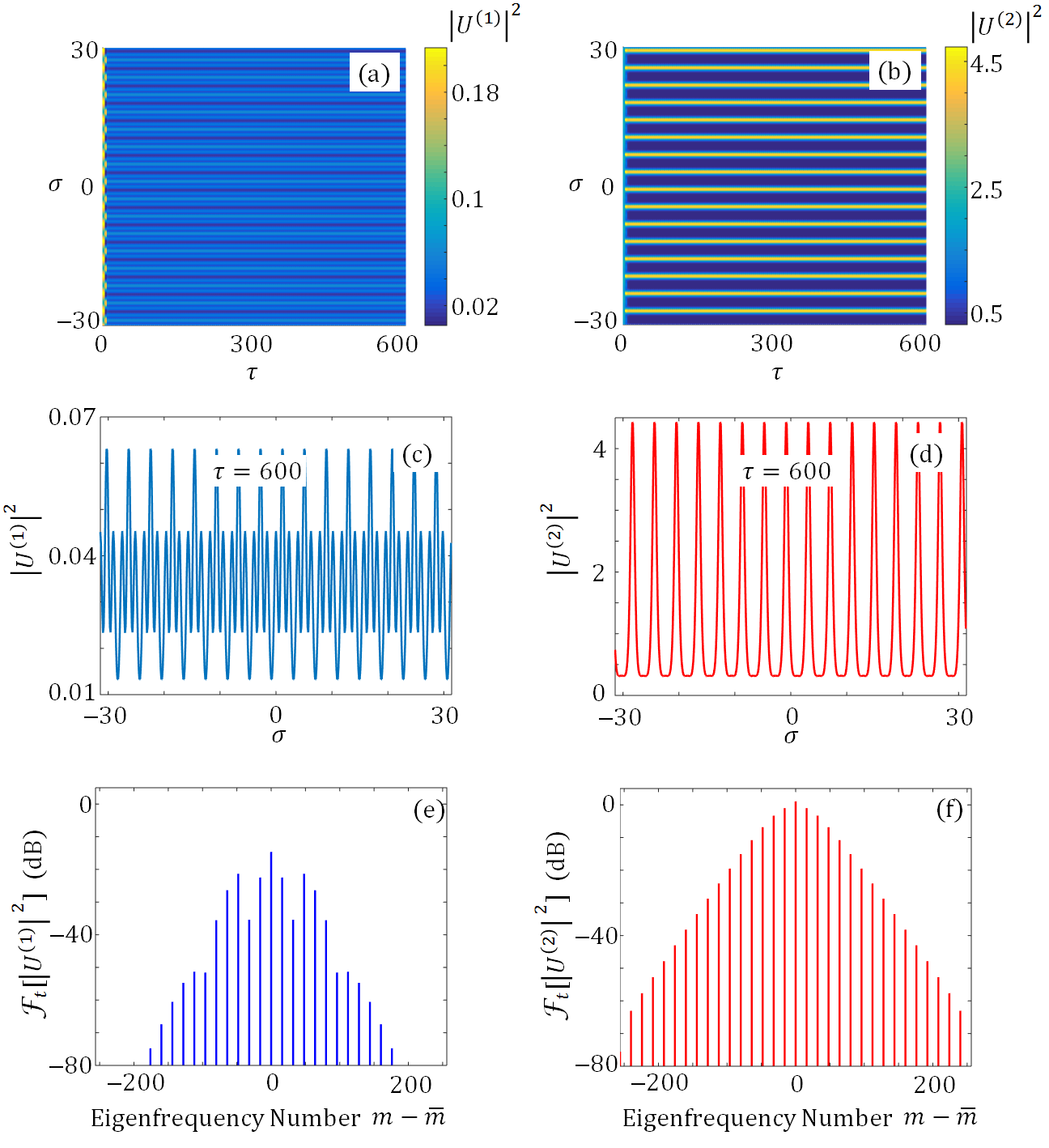}}
\end {center}
\caption{ Example of hyperparametric oscillations calculated for $P^{(1)}=0.15$ and $P^{(2)}=1.95$. (a) and (b):  Spatio-temporal evolution of the numerical solutions. (c) and (d): Numerical solutions calculated at $\tau=600$. (e) and (f): Fourier transform (${\mathcal{F}}_{t}$) of the solutions.}
\end{figure*}
In the Fourier space, these hyperparametric oscillations are characterized by a coarse-tooth frequency comb. Although for many applications a dense-tooth frequency comb is generally more desirable than a coarse-tooth one, that is not always the case. Frequency combs with coarse-tooth characteristics are useful, for example, in quantum networking \cite{Roslund} or astrocombs \cite{Li}, where limiting the number of comb lines and precisely controlling their amplitudes is required. In the transition region between the periodic patterns and the hyper-parametric oscillations, the field initially self-organizes into periodic patterns, but then at some point suddenly changes into a  CW solution.   Finally, by further increasing the pump level on the anomalous dispersion branch, ($P^{(2)}\gg P^{(1)}$), solitons, and hence dense-tooth (broadband) frequency combs, become accessible.  In Fig.~5, we show an example of a 2-soliton state. To exploit this mechanism, two different pumps are needed, where each of them is closely detuned from only one of the transverse modes.\\
\begin {figure*}
\begin {center}
\resizebox{1\textwidth}{!}{%
  \includegraphics {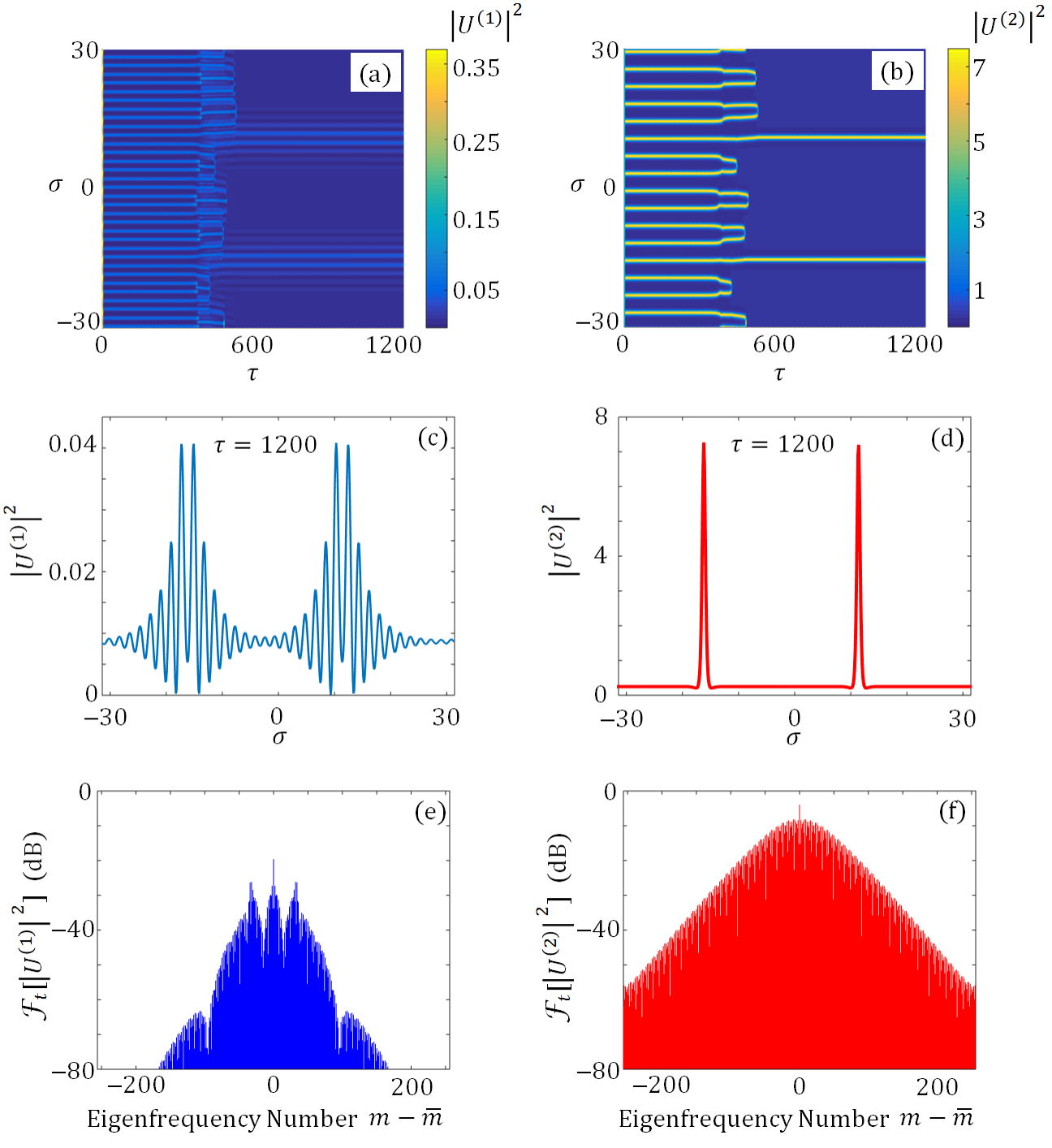}}
\end {center}
\caption{Example of a 2-soliton state calculated for $P^{(1)}=0.3$ and $P^{(2)}=3$. (a) and (b):  Spatio-temporal evolution of the numerical solutions. (c) and (d): Numerical solutions calculated at $\tau=1200$. (e) and (f): Fourier transform (${\mathcal{F}}_{t}$) of the solutions.}
\end{figure*}

\indent We emphasize two points. First, the soliton generation just described  strictly depends on the presence of the cross-coupling term. Hence, it is the result of the nonlinear interaction among the two hybrid modes. If the cross-coupling term in Eq.~(3) is neglected, solitons are not generated. Second, the region of hyperparametric oscillations is directly connected to the region of bright solitons with no chaotic region in between. Hence, we have identified a deterministic path for soliton generation. By contrast, in standard 
soliton generation in microresonators \cite{Herr}, a region of chaotic oscillations separates the region of  solitons from the region of hyperparametric oscillations.  Our finding is consistent with previous experimental results where, close to an avoided crossing of a microresonator, the direct generation of coherent, bandwidth-limited pulses (solitons) has been observed without the need to first pass through a chaotic state \cite {Weiner}. 
\pagebreak
\section {Conclusions}
\indent 
 In conclusion, we have studied nonlinear  mode coupling in WGM resonators at an avoided crossing, and we have found a deterministic path to generate bright solitons and broadband combs. The generation of bright solitons in a microresonator is often believed to only be possible in the anomalous dispersion regime. Yet, it would be highly beneficial for many applications to generate bright solitons and broadband combs in the visible and near-UV, where most of the dielectric materials have normal dispersion. Recently, several approaches have been proposed to broaden the frequency comb into the normal dispersion region. Such approaches include, among others, the generation of Cherenkov radiation \cite {Brasch,Matsko1}, second harmonic generation \cite {Leo,Xue}, and the use of concentric-racetrack-resonators \cite{Kim}.  In this work, we have proposed a different approach that is based on the peculiar dispersion properties that can be achieved at an avoided crossing.  We have shown that bright solitons and broadband combs can be deterministically generated at an avoided crossing by using two independent pumps at different pump frequencies. Regardless of whether the resonator is in a normal or an anomalous dispersion region, an avoided crossing provides two branches with dispersion of opposite sign  whose nonlinear coupling can lead to a deterministic path for bright soliton and broadband comb generation. \\
\section {Acknowledgments}
This work was supported in part by ARL Project No. W911NF-13-2-0010 and by AMRDEC/DARPA Project No. W31P4Q-14-1-0002. The numerical simulations
were carried out at UMBC's high performance computing facility. We thank Andrew Weiner, Minghao Qi, Xiaoxiao Xue, Jose Jaramillo-Villegas, Zhen Qi, Thomas Carruthers, and Andrey Matsko for useful discussions.
\section {Author contribution statement}
G.D. developed the analytical model, performed the numerical simulations and wrote the paper. C.R.M contributed to the writing of the paper and the interpretation of the results.
%
%
%

\end{document}